\documentstyle[12pt,psfig]{article}
\begin{document}
\begin{center}
{\bf Size-selective concentration of chondrules and other small particles\\
in protoplanetary nebula turbulence}\\

Jeffrey N. Cuzzi$^1$, Robert C. Hogan$^2$, Julie M. Paque$^3$, and Anthony R. 
Dobrovolskis$^4$\\

\begin{footnotesize}
1) Ames Research Center, NASA, cuzzi@cosmic.arc.nasa.gov; 2) Symtech, inc.,
hogan@cosmic.arc.nasa.gov; 3) SETI Institute, jpaque@mail.arc.nasa.gov; 4)
Univ. of California, Santa Cruz, dobro@cosmic.arc.nasa.gov; mailing address for 
all authors: Mail Stop 245-3, Moffett Field CA 94035-1000.
\end{footnotesize}

Astrophysical Journal, in press (expect January 1 2001 issue)

\end{center}

\begin{abstract}
Size-selective concentration of particles in a weakly turbulent protoplanetary
nebula may be responsible for the initial collection of chondrules and other
constituents into primitive body precursors. This paper presents the main
elements of this process of {\it turbulent concentration}. In the terrestrial
planet region, both the characteristic size {\it and} size distribution of
chondrules are explained.  ``Fluffier" particles would be concentrated in
nebula regions which were at a lower gas density and/or more intensely turbulent.
The spatial distribution of concentrated particle density obeys {\it
multifractal scaling}, suggesting a close tie to the turbulent cascade process.
This scaling behavior allows predictions of the probability distributions for
concentration in the protoplanetary nebula to be made. Large concentration
factors ($>10^5$) are readily obtained, implying that numerous zones of
particle density significantly exceeding the gas density could exist. If most
of the available solids were actually in chondrule sized particles, the ensuing
particle mass density would become so large that the feedback effects on gas
turbulence due to mass loading could no longer be neglected. This paper
describes the process, presenting its basic elements and some implications,
without including the effects of mass loading.

\end{abstract}

\section{Background and Introduction:} Primitive (unmelted) chondritic
meteorites are composed in large part of mm-sized, once-molten silicate
particles (chondrules) and metallic grains out of mineralogical equilibrium
with each other. Many chondrites contain inclusions of refractory minerals that
have been dated as the oldest objects formed in the solar system (MacPherson et
al 1989, 1995). The chemical, isotopic, mineralogical and petrographic
properties of individual chondrules themselves imply that independent entities
were melted by some ``flash heating" event in the gaseous protoplanetary
nebula, and remained molten for fairly short times (less than an hour; Jones et
al 2000). Chondrules are diverse in chemistry but are narrowly size-sorted
(Grossman et al 1989, Jones and Brearley 1999), apparently by their aerodynamic
cross section (Dodd 1976, Skinner and Leenhouts 1991, Keubler et al 1999); the
least mechanically evolved pieces in chondrites have the appearance of being
gently brought together (Metzler et al 1992) and subsequently compacted to
solid density. Various hypotheses have been advanced to explain these
properties (see, e.g., Boss 1996, Hewins et al. 1996, Hewins 1997, Connolly and
Love 1998, and Jones et al 2000 for recent general reviews and discussion), but
the puzzle remains unsolved. The fact that many primitive meteorites are
composed of up to 70-80\% chondrules by volume (Grossman et al 1989) implies
that the chondrule formation and accumulation processes were of widespread and
significant importance in the very earliest stages of the accretion of the
asteroidal objects which provide the parent bodies for these primitive
meteorites. Since the terrestrial planets apparently formed from various
proportions of known meteorite types, one expects that the process extended
beyond the current asteroid belt and that understanding these early stages is
important for understanding planetary accretion overall. These stages of
circumstellar disk evolution relates to regions and epochs when the particulate
opacity is high, but when some particle growth has occurred - and are thus also
of great interest for study at infrared and millimeter wavelengths.

Prior work (Dubrulle et al 1995, Cuzzi et al 1996) has pointed out that, unless
the turbulent kinetic energy in the nebula gas is vanishingly small (see {\it
Section 2}), chondrule-sized particles are unable to settle individually to the
nebula midplane, where most growth to planetesimal size must occur. Instead,
Cuzzi et al (1996; henceforth CDH96) proposed that, following their initial
melting, and throughout multiple recurrences of similar heating events (Wasson
1996, Connolly and Love 1998, Desch and Cuzzi 2000), chondrules pursue an
extended free-floating existence under plausible conditions of nebula gas
density and turbulent intensity in the terrestrial planet region, successively
encountering zones of varying concentration enhancement (section 5.1), until by
chance they encounter an unusually dense zone where they might physically
coalesce into much more massive, but still not solid, entities. In a second
stage, such entities - dense clusters of particles - might have enough
coherence to resist disruption as they settle to the midplane, or to be
collected in a different process into the cores of the largest eddies
(discussed below) for subsequent accumulation into planetesimals. Or, dense
zones may only provide environments of enhanced collisional accumulation of
chondrules. The second stage remains unstudied and qualitative; here we focus
on the first stage.

The process of interest is known as {\it preferential concentration} or {\it
turbulent concentration} (TC). The nature of this process is that isotropic,
homogeneous, 3D turbulence provides numerous fluid zones of low vorticity and high
strain in which particles having a narrowly defined range of aerodynamic
properties can be significantly, but usually briefly, concentrated (discussed
further below). This somewhat counterintuitive effect was first alluded to
theoretically (Maxey 1987), subsequently demonstrated numerically (Squires and
Eaton 1990, 1991; Wang and Maxey 1993), and recently demonstrated
experimentally as well (Fessler et al 1994; for a review see Eaton and Fessler
1994). The {\it concentration factor} $C$ is the ratio of the local particle
density to its global average. In numerical studies to date (Squires and Eaton
1990, 1991; CDH96, Hogan et al 1999), turbulence with Reynolds numbers of $10^2
- 10^3$ contains dense zones where the concentration factors $C$ reach 40-300
for optimally concentrated particles. We extended numerical studies of this
effect, and applied scaling relationships to predict its behavior under
protoplanetary nebula conditions (CDH96, Cuzzi et al 1998). We found that under
canonical inner nebula conditions, particles with the size and density of
chondrules would be optimally concentrated. Here, we show that the process is
easily generalized to a wide range of fluffier particles in lower density
regions of the nebula ({\it section 3}).

We note that TC is a quite different process than the superficially similar
effect in which far larger (meter-and-larger-radius) particles with stopping
times comparable to or larger than the eddy times of the {\it largest} eddies
(with frequencies comparable to the orbit period) can accumulate near the
centers of such eddies (Barge and Sommeria 1995, Tanga et al 1996, Bracco et al
1999). Klahr and Henning (1997) showed concentration of mm-sized particles
within large, slow, 2D circulation patterns (not turbulent eddies). However,
unless these slow circulation patterns represent the primary kinetic
energy reservoir of the (non-Keplerian) fluid motions, concentration of such 
small particles will not occur. That is, if the nebula turbulence has a normal 
3D cascade with its energy peak at spatial and temporal scales which are a
small fraction of the nebula scale height and orbital period (as normally
implied by turbulent ``$\alpha$-models"described in section 2),
chondrule-sized particles will diffuse faster than they can be concentrated by
such large, slow circulation patterns. For example,
numerical calculations by Supulver and Lin (1999), which modeled eddy
motions on a wide ``inertial range" of spatial and temporal scales, as well as
our own calculations, show that chondrule-sized particles do not accumulate in
large eddies for two reasons: first, their trajectories are mixed by the
smaller eddies, and second, ``real" eddies in homogeneous turbulence - even the
largest ones - dissipate within a single overturn time. Under these
circumstances (assumed in this paper), mm-sized particles would be nearly
uniformly dispersed by 3D turbulence, as one would naively expect for such
small particles which are trapped to nearly all fluid parcels, if it were not
for the role of turbulent concentration as described herein. The true energy
spectrum and dimensionality of nebula turbulence is, however, not currently
understood, and is the subject of several active research tasks.

Our emphasis in testing these concepts has been to compare our predictions with
the properties of easily identified and studied chondrules and chondrites. For
instance, we have now studied the size {\it distribution} of preferentially
concentrated particles in detail, and found it to be insensitive to, or
independent of, Reynolds number (Hogan and Cuzzi 2000); here we show that this
predicted size distribution is in very good agreement with a typical chondrule
size distribution ({\it section 4}).

While $C$ does increase systematically with increasing Reynolds number (CDH96),
our prior estimates of concentrations at nebula Reynolds numbers, which are
plausibly far larger than those accessible to numerical modeling, had required
sizeable extrapolations. We have more recently shown that the spatial structure
of the concentrated particle density field is a multifractal which has
Reynolds-number-independent properties (Hogan et al 1999) and here will use
this result to provide a firmer basis for predictions under nebula conditions
({\it section 5}).

The evolution of extremely dense clumps, within which interparticle collisons
might entrap particles  (CDH96) remains unstudied, and will require a better
understanding of the behavior of turbulent concentration when the particle mass
density exceeds that of the gas, and of particle ensembles whose density is
large enough to affect the gas flow properties. In {\it section 6} we discuss 
this important effect.

\section{Turbulence and turbulent concentration (TC):} Homogeneous, isotropic,
3D turbulence is characterized by a cascade of energy through a range of
scales, known as the inertial range, from the largest (or integral) spatial
scale $L$, having associated velocity $V_L$, to the smallest (or Kolmogorov)
scale $\eta$ where it is dissipated (Tennekes and Lumley 1972, Hinze 1975). 
The intensity of the turbulence is characterized by the Reynolds number, which 
can be written $Re \equiv (L/\eta)^{4/3}$. Since dissipation occurs primarily
on the small scales where molecular viscosity comes into play, more energetic
(higher $Re$) flows can drive turbulence through a wider inertial range, to
smaller $\eta$, for any given viscosity.

$Re$ is ordinarily defined as $Re=LV_L/\nu_m$, where $\nu_m$ is the molecular
viscosity; $Re$ here is the ratio of transport by macroscopic motions to that
by molecular motions. This definition combines the velocity and length scales
into a {\it turbulent viscosity} $\nu_T$. The Reynolds stresses which appear in
angular momentum transport equations are often modeled by this sort of
turbulent viscosity (the ``alpha model" of Shakura and Sunyaev 1973). However,
in a Keplerian disk, the mere existence of turbulent motions (turbulent kinetic
energy, leading to diffusivity of scalars) does not necessarily imply viscous
transport of angular momentum (by a Reynolds stress, or torque, acting as a
positive turbulent viscosity; Prinn 1990). This distinction is related to the
possibility that the on-diagonal and off-diagonal terms of the stress tensor
might have very different relative strengths in Keplerian disks than in more
familiar turbulent environments (Kato and Yoshizawa 1997). While astrophysical
``alpha-models" of protoplanetary disks emphasize turbulent viscosity $\nu_T$
in its angular momentum transport role, TC is more closely related to scalar
diffusivity, or turbulent kinetic energy per unit mass density $k$. Therefore,
we distinguish two types of the familiar astrophysical $\alpha$. We first
discuss the familiar Shakura-Sunyaev prescription, which defines the turbulent
viscosity as $\nu_T = \alpha_{\nu} c H$, where $c$ is the sound speed and $H$
is the vertical scale height.

Without external drivers, the overall nebula is likely to be in a regime of 
Rossby number $Ro \equiv \Omega_L/\Omega_0$ of order unity, where $\Omega_L$ is
the largest eddy frequency and $\Omega_0$ is the orbital frequency. We thus
rewrite $\nu_T$ in terms of these fundamental properties:
\begin{equation}
\nu_T = L V_L = L^2 \Omega_L \hspace{0.1 in} (= V_L^2/\Omega_L) = \alpha_{\nu}
c H = \alpha_{\nu} H^2 \Omega_0 \hspace{0.1 in} (= \alpha_{\nu} c^2/\Omega_0);
\end{equation}
Thus 
\begin{equation}
L = H \sqrt{\alpha_{\nu}}\left( { \Omega_0 \over \Omega_L} \right)^{1/2},
\hspace{0.1 in} 
V_L = c \sqrt{\alpha_{\nu}}\left( { \Omega_L \over \Omega_0} \right)^{1/2}.
\end{equation}
With the stipulation that the largest {\it truly turbulent} eddies (those that 
participate in the turbulent cascade) have frequencies $\Omega_L$ no smaller
than, but probably comparable to, the local orbital frequency $\Omega_0$, $V_L
= c \sqrt{\alpha_{\nu}}$ and $L = H \sqrt{\alpha_{\nu}}$. Occasionally it is
assumed that $L \approx H$, but this implies $V_L = \nu_T/L = c \alpha_{\nu}$
and thus $\Omega_L = V_L/L = c \alpha_{\nu}/H = \alpha_{\nu} \Omega_0 <<
\Omega_0$, an implausible situation for genuine turbulence. The distinction is
important for us, as we need to scale turbulent velocities and lengthscales
independently. The important point is that the scaling parameter $\alpha_{\nu}$
is ``shared" by the length and velocity scales, rather than being associated
with one or the other.

A scaling analysis cannot establish whether fluid motions with arbitrary $L$ 
and $V_L$ do in fact provide a positive Reynolds stress or ``turbulent viscosity"
given by their product; conversely, estimates of the magnitude of $V_L$ and $L$ 
from an observed turbulent viscosity $\nu_T$ and associated $\alpha_{\nu}$, as
above, may misrepresent their actual magnitudes. This is because certain types
of spatial correlations between ``random" fluid motions are needed to provide a
positive Reynolds stress, or turbulent viscosity, and this may not occur in
systems which are strongly influenced by rotation or otherwise strongly
perturbed (Prinn 1990, Kato and Yoshizawa 1997).

Another approach to determining $V_L$ is based on the turbulent kinetic energy
per unit mass $k$ and an associated $\alpha_k$: $k = \frac{3}{2} V_L^2 \equiv
\frac{1}{2}\alpha_k c^2$, or $V_L= c\sqrt{\alpha_k/3}$. The mere presence of
turbulent fluid motions with $k \equiv \alpha_k c^2/2 = 3 V_L^2/2$ is sufficient
to produce turbulent concentration, with no additional uncertainties about
the degree and sign of the correlation between orthogonal components of the 
fluid motions as in $\alpha_{\nu}$ (Prinn 1990). While this definition provides
no insight into the turbulent length scale $L$, we presume by analogy to the
above argument that $L = H \sqrt{\alpha_k}$. With regard to the nebula, where
properties are uncertain, we suppress the factor of $\sqrt{3}$ (effectively
ignoring the distinction between a single component of $V_L$ and its magnitude)
and approximate the turbulent Reynolds number of the nebula by $Re = \alpha_k c
H/\nu_m$. Angular momentum may still be transported by a turbulent viscosity
$\nu_T = \alpha_{\nu} c H$, but whether $\alpha_{\nu} \approx \alpha_k$ or not
is peripheral to this work (Kato and Yoshizawa (1997) find that $\alpha_{\nu} <
\alpha_k$).

Possible sources of $k$ include forcing by ongoing infall onto the disk in the
very early stages (Cameron 1978, Prinn 1990), turbulent convection powered by
release of gravitational energy into heat as the disk evolves (Lin and
Papaloizou 1985, Cabot et al 1987; Goldman and Wandel 1994; Bell et al 1997),
enforced Keplerian differential rotation (Dubrulle 1993), and magnetorotational
instability or MRI (Balbus and Hawley 1991, 1998). Typical estimates of
$\alpha_k$ from these sources are $\sim 10^{-4}$ - $10^{-1}$, and thus $Re =
10^8$ - $10^{11}$.

However, uncertainties remain with all of the above mechanisms. The infall
stage lasts only a relatively short fraction of the evolution lifetime of
typical nebulae, and the magnetorotational instability will not act in the
dense inner scale height of the disk where most of the mass and chondrules
probably reside (Gammie 1996) and perhaps not even high in the nebula (Desch
2000). The validity of convection and differential rotation, acting by
themselves, rests on their uncertain ability to be self-sustaining. This
requires turbulence to be able to transport sufficient angular momentum
outwards that mass can evolve inwards, releasing gravitational energy to be
converted into turbulence. Several studies (most recently Stone and Balbus
1996) found that convective turbulence fails to produce outward angular
momentum transport (positive Reynolds stresses). However, some recent 3D
numerical studies, which better capture large azimuthal structures, seem to
imply that it may in fact be able to do so (Klahr 2000a,b). Of course, as grains
accumulate, opacity decreases and thermal convection may weaken. Differential
rotation is free of this limitation (Dubrulle 1993), but numerical and
analytical arguments by Balbus et al (1996) question differential rotation as a
source for turbulence based on the energetics and stability of Keplerian disks.
However, Richard and Zahn (1999) have suggested, based on laboratory analogues,
that instability to turbulence in such systems requires a higher Reynolds
number than accessible to current numerical models. Kato and Yoshizawa (1997)
have shown that Keplerian rotation probably does not preclude some true
turbulent viscosity ({\it ie.} positive Reynolds stress, or $\alpha_{\nu} >
0$), but as a much smaller fraction of the ambient turbulent kinetic energy
($\alpha_k$) than under non-Keplerian conditions. Thus, it also remains in
doubt whether turbulence driven by differential rotation alone can be
self-sustaining. Non-steady situations might even need to be considered.

In short, the source of angular momentum transport is poorly understood, and it
is not clear how disks evolve at all. However, while we still don't understand
the {\it mechanism} that allows them to do so, protoplanetary disks are {\it
observed} to evolve with mass accretion rates in the range of $\dot{M} =
10^{-7} - 10^{-9} M_{\odot}$/yr (Hartmann et al 1998) and thus have an
associated gravitational energy release rate per unit area of $\dot{E}_{\rm
grav} = 3 G M_{*}\dot{M}/4 \pi R^3$ (Lynden-Bell and Pringle 1974). Here, $G$
is the gravitational constant, $M_{*}$ is the mass of the central star, and $R$
is the distance from the central star.

Turbulent kinetic energy is dissipated at a rate of approximately $\dot{E}_k= 2
k \Omega_0  \rho_g H = k \Omega_0 \Sigma$, where $\rho_g$ is the local gas
density and $\Sigma = 2 \rho_g H$. If the gravitational energy $\dot{E}_{\rm
grav}$ is released where most of the mass resides, and converted into
mechanical turbulence with efficiency $\xi_T$, then for a disk in steady state
with $\dot{M}$ independent of radius, $\dot{E}_{\rm grav}\xi_T  = \dot{E}_k$, or
\begin{equation}
{3 G M_* \dot{M} \xi_T \over 4 \pi R^3} = k \Omega_0  \Sigma
= {\alpha_k c^2\Omega_0 \Sigma \over 2}. \\
\end{equation}
Substituting $ \Omega_0^2 = G M_*/R^3$ and $c=H \Omega_0$, 
\begin{equation} 
\alpha_k = {3 \dot{M} \xi_T \over  2 \pi c H \Sigma}.
\end{equation}
For a typical ``canonical" nebula with $H(R)=0.05 {\rm AU} (R/1{\rm
AU})^{5/4}$, $c(R)=1.9 \times 10^5(R/1{\rm AU})^{-1/4}$ cm sec$^{-1}$, and
$\Sigma(R)=1700{\cal F}(R/1{\rm AU})^{-3/2}$g cm$^{-2}$, with ${\cal F}$ being
any enhancement in mass over the ``minimum mass" nebula density (Hayashi 1981; 
see also Cuzzi et al 1993), then
\begin{equation}
\alpha_k \approx  1.3 \times 10^{-3} \left({R \over 1{\rm AU}}\right)^{1/2} 
{\dot{M} \over 10^{-8} M_{\odot}/{\rm yr}} {\xi_T \over {\cal F}}.
\end{equation}
Some recent numerical calculations show $\xi_T$ to be at least several percent
(H. Klahr, personal communication 1999); thus $\alpha_k \sim 10^{-3} - 10^{-4}$
seems not to be out of the question in the 2-3 AU region where meteorite parent
bodies form. Conversely, Dubrulle et al (1995) have shown that, in order for
chondrule-sized particles to settle into a midplane layer having a density 
approaching that of the gas (a layer of thickness $10^{-2} - 10^{-3}H$),
$\alpha_k$ would need to be in the $10^{-8} - 10^{-10}$ range (see also CDH96).
For $\alpha_k$ to be this low, equations 4 and 5 show that the conversion
efficiency into turbulence would need to be less than $10^{-5}$ for a disk with
${\dot{M} = 10^{-8} M_{\odot}/{\rm yr}}$. Here we simply presume the presence of
nebula turbulence at a weak level and explore the consequences. Henceforth, we
identify the nebula ``$\alpha$" as $\alpha_k$, and treat $Re$ and its
associated $\alpha$ as characterized by $L$ and $V_L$. 

We assume the turbulence has a Kolmogorov-type inertial range, within which
each length scale $l$ is characterized by velocity $v(l) = V_L (l/L)^{1/3}$
(Tennekes and Lumley 1972, CDH96) and eddy frequency $\omega(l) = v(l)/l =
\Omega_0 (l/L)^{-2/3}$, where the frequency of the largest eddy $\omega(L) =
V_L/L$ is set equal to $\Omega_0$, the local orbital frequency. Because the
most interesting scales for particle concentrations are on the order of $\eta =
L Re^{-3/4} << L$, and even $L \approx H \sqrt{\alpha} << H$, deviations from
isotropy due to rotation are not a major concern. Particles smaller than the
gas mean free path ({\it i.e.}, smaller than several cm radius under nebula
conditions) are in the Epstein drag regime (Weidenschilling 1977), and have a
stopping time due to gas drag which is
\begin{equation}
t_s = r \rho_s / c \rho_g,
\end{equation}
where $r$ and $\rho_s$ are particle radius and internal density. The particle
Stokes number $St_l \equiv t_s \omega(l)$ determines the particle response to
eddies of a particular size and frequency; previous studies have shown that
the optimally concentrated particles have $St_{\eta}= t_s \omega(\eta) \approx
1$ (Eaton and Fessler 1994; Hogan and Cuzzi 2000), that is, their stopping time
is comparable to the Kolmogorov eddy turnover time.

\section{Generality of turbulent concentration}
Solving the relation  $St_{\eta} = t_s \omega(\eta) = 1$ under nebula
conditions, CDH96
concluded that the optimally concentrated particles in the terrestrial planet
region of a minimum mass nebula with $\alpha \sim 10^{-4}-10^{-3}$ 
would have radius and density comparable to those of
chondrules. Here we generalize TC to the entire 
range of nebula
conditions. Using the definition for $t_s$ (equation 6) to rewrite 
the expression for $St_{\eta}= t_s \omega(\eta) =1$,
\begin{equation}
r \rho_s = c \rho_g t_s = c \rho_g t_{\eta} 
= { c \rho_g \over \Omega_0(\eta/L)^{-2/3}} 
= {c \rho_g \over  \Omega_0 Re^{1/2}}
= {c \rho_g \nu_m^{1/2} \over \Omega_0 (\alpha c H)^{1/2}}.
\end{equation}
CDH96 Substituted $\nu_m = c \lambda/2 = m_{H_2} c / 2 \rho_g \sigma_{H_2}$,
where $\lambda$ is the molecular mean free path, and $m_{H_2}=3.2\times
10^{-24}$g and $\sigma_{H_2}=5.7\times 10^{-16}$cm$^2$ are the mass and cross
section of a hydrogen molecule. However, their expression for $\lambda$ did not
include the finite size or the Maxwellian velocity distribution of the gas
molecules ({\it cf.} Kennard 1938; equations 106d and 126b), so their $\lambda$
was too large by a factor of $4\sqrt{2}$. Rearranging terms from equation (7)
and correcting this oversight, we find
\begin{equation}
r \rho_s = \left({m_{H_2} \over 16 \sqrt{2} \sigma_{H_2}}\right)^{1/2}
\left({\Sigma \over \alpha}\right)^{1/2}
= 6.3 \times 10^{-4} \left({{\cal F} \over \alpha}\right)^{1/2} 
\left({R \over {\rm 1AU}}\right)^{-3/4} {\rm g}\hspace{0.02in}{\rm cm}^{-2},
\end{equation}
where $\Sigma(R)= 2 \rho_g H$ is surface mass density at some distance in the
nebula. In the last term we have adopted a canonical radial dependence of
$\Sigma(R)= 1700{\cal F}(R/{\rm 1 AU})^{-3/2}$ for a ``minimum mass" nebula,
with ${\cal F}$ being some mass enhancement factor.

\begin{figure}
\centerline{\psfig{figure=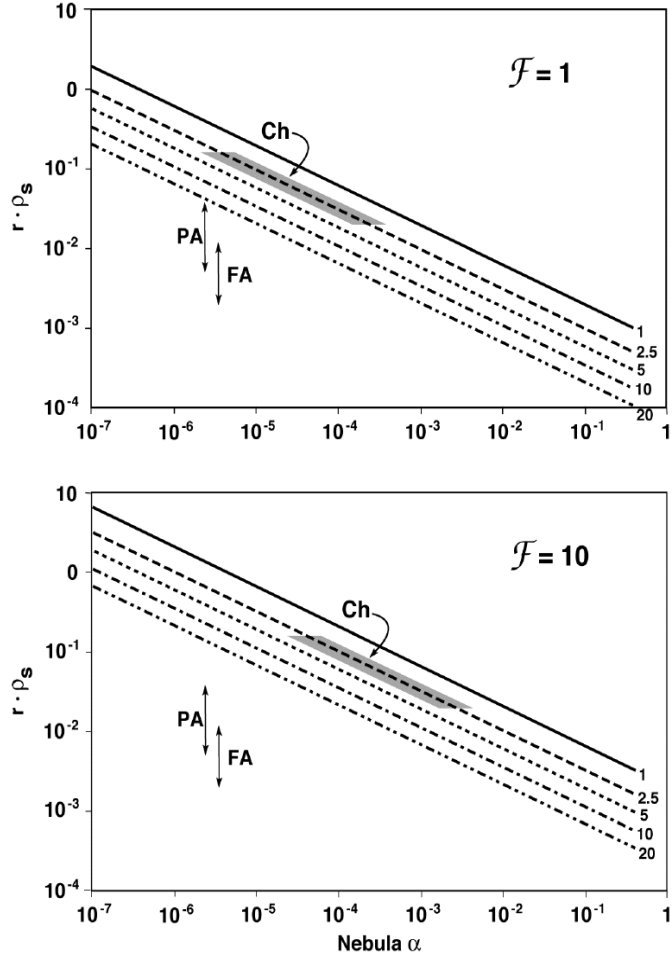,width=3.5 in,height=5.0 in}}
\caption{The optimal particle radius-density product $r \rho_s$ for turbulent
concentration as a function of $\alpha$ at various locations in a typical solar
nebula (converted from the optimal Stokes number $St_{\eta}=1$ using nominal
nebula gas density and local rotation rate). (a) Minimum mass nebula (${\cal F} 
=1$); Solid silicate particles with chondrule sizes and densities (Ch) have the
range $r \rho_s \approx 0.015 - 0.15$, which would indicate the shaded range of
$\alpha$ at 2.5 AU. Smaller gas densities at larger distances from the sun
concentrate smaller $r \rho_s$ products (porous aggregates or PA) and even
``fluffier" $d$=2 fractal aggregates (FA) or their monomers, for any $\alpha$.
(b) Somewhat different results obtained assuming an enhanced surface mass
density ${\cal F} = 10$ times higher than the minimum mass requirement. }
\end{figure}

This relationship is shown in {\bf figure 1} for several different typical
locations and two different nebula masses. The range of $r \rho_s$ for
most chondrules (Ch), obtained using data in Grossman et al (1989), 
is mapped along the line for $R=2.5$ AU, and indicates the
range of $\alpha$ required to concentrate them selectively. Higher values of 
$\alpha$ (more intense turbulence) select smaller and/or lower density 
particles. In both cases,
porous aggregates (PA), having considerably lower radius-density product than
chondrules, can be optimally concentrated at the low gas densities which
characterize the outer planet region or low density regions high above the
nebula midplane. Such porous objects are easily produced because the low
relative velocities in turbulence of both porous, low-density aggregates, and
their constituent monomers, lead to large sticking efficiency and minimal
disruption (Weidenschilling and Cuzzi 1993; Chokshi et al 1993; Dominik and 
Tielens 1997). It has also been suggested that extremely low density,
intensely turbulent regions may concentrate tiny grains or very fluffy
aggregates (FA) (Wood 1998). TC may thus have been ubiquitous, helping to 
initiate the formation of ``cometesimals" from porous grain aggregates at 10-30
AU or speeding the accretion of dust particles at high elevations (low gas
densities) even in the terrestrial planet region.

However, textural evidence might be difficult to obtain from this regime;
subsequent compaction would obliterate evidence for any preferred size or
density of easily squashed fluffy constituents. Chondrules and their parent
chondrites, by nature of their availability and unique, persistent textures,
provide the most obvious initial testing ground for turbulent concentration.

\section{Concentrated particle size distribution}

If dense clusters of particles are precursors of primitive bodies in any way,
the relative abundance as a function of Stokes number $St_{\eta}$ for particles
in dense regions should show some correspondence to the distributions found in
chondrites. To further compare the predictions of TC with meteorite evidence,
we have recently determined the detailed form of the size distribution of
selectively concentrated particles (Hogan and Cuzzi 2000). Numerical
simulations were performed with homogeneous, isotropic, incompressible 3D
turbulence at three Reynolds numbers $Re =$ 62, 246, and 765 (these values 
differ from those given in Hogan et al. 1999 because we now adopt the more 
correct definition of $V_L$ as one component of the turbulent velocity rather 
than its magnitude). The particles can be given arbitrary aerodynamic stopping
times $t_s$; their motions respond only to gas drag and are integrated in the
spatial domain. Feedback by the particles onto the gas is not incorporated. The
computationally intensive calculations are run on 16 Cray C90 cpus at Ames
Research Center. Simulations were initiated with uniform spatial distributions
of particles, themselves uniformly distributed in stopping time over the range
$St_{\eta}$ = 0.1 - 6, and the relative equilibrium abundance of particles was
studied as a function of $St_{\eta}$ and $C$. In the large-$C$ limit of
interest, the shape of the relative abundance distribution ${\cal
A}(St_{\eta})$ was found to be {\it independent} of both $C$ and $Re$ and only
very weakly dependent on the spatial binning assumed (relative to $\eta$)
(Hogan and Cuzzi 2000). Thus, we believe the numerical results should be valid
as a prediction of the size distribution in dense particle concentrations under
nebula conditions. The theoretical results are adequately fit by a lognormal
distribution over the core range $St_{\eta}$ = 0.5 to 2.0.

\begin{figure}
\centerline{\psfig{figure=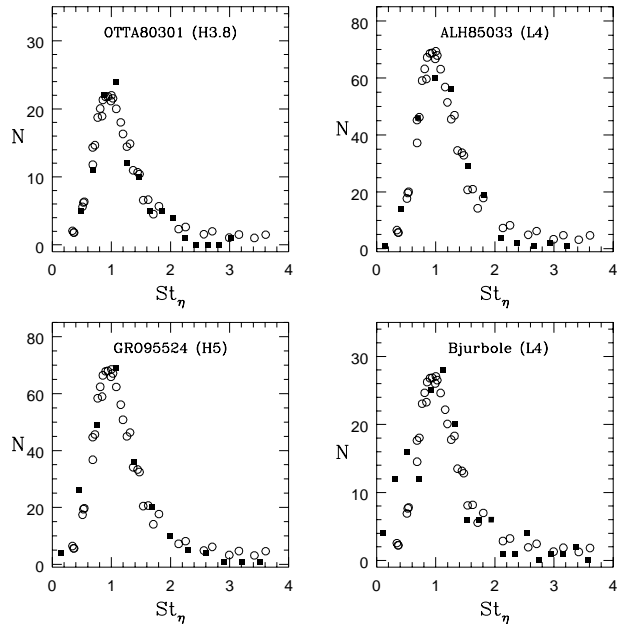,width=3.5 in,height=3.5 in}} 
\caption {Comparison of the size-density distributions of chondrules
disaggregated from four ordinary chondrites (solid symbols; Paque and Cuzzi
2000) with the theoretically determined, $Re$-independent, shape distribution
for the relative abundance of particles concentrated in turbulence as a
function of $St_{\eta}$ (open symbols; Hogan and Cuzzi 2000).}
\vspace{0.5 in}
\end{figure}

We compared our predictions with binned relative abundance data for 
chondrules disaggregated from one carbonaceous and four ordinary chondrites
(Paque and Cuzzi 1997, 2000; Cuzzi et al 1999). While a few ostensibly similar
data sets exist in the literature ({\it eg}. Hughes 1978, Eisenhour 1996, Rubin
and Grossman 1987), they generally rely on radii measured microscopically from
thin sections of meteorites, or even from disaggregated chondrules, and adopt
some average chondrule density $\overline{\rho_s}$. However, recall from
equation (6) that the aerodynamic stopping time $t_s$, which selects particles
for TC or any other aerodynamic sorting process, depends on the product $r
\rho_s$ for {\it each} object. Data for these meteorites imply that merely
measuring the chondrule radius distribution and assuming some mean density
will slightly, but noticeably, misrepresent the $r \rho_s$ distribution because
of chondrule-to-chondrule density variations (Cuzzi et al 1999, Paque and Cuzzi
2000).

In {\bf figure 2} we compare the particle abundance shape function ${\cal
A}(St)$ from our numerical simulations (Hogan and Cuzzi 2000) with the
meteorite data. The meteorite histograms were fit by lognormal functions and
aligned horizontally with the predicted histogram by assuming that the
optimally concentrated particle size-density product corresponds to
$St_{\eta}=1$. We conclude that TC by itself can explain the very narrow
chondrule size distribution, whatever the chondrule {\it formation} process may
have produced. There is evidence that both larger and smaller chondrules were
created; TC would predict that these non-optimally sized particles were simply
not concentrated to a sufficient degree for eventual incorporation into 
meteorites unless they were swept up into chondrule rims (if originally
smaller) or broken to the proper size (if originally larger). Some particles of
arbitrary size will always appear just by accident, perhaps captured by a dense
cluster; parent body processes will confuse the situation further.

\section{Probability distributions and multifractals:} As mentioned earlier,
our numerical results (CDH96) are obtained at $Re$ which are six orders
of magnitude smaller than plausible nebula $Re$. Thus, it is very important to
understand the $Re$-scaling of TC properties in detail. We believe this has
become possible using a connection between TC and the $Re$-independent scaling
properties of fractals, which are rooted in the properties of the turbulent
cascade itself (Hogan et al 1999; henceforth HCD99). 
Fractals and multifractals are readily associated with {\it
cascade processes}. Turbulence (and specifically its inertial range)
is the archetype of a cascade process (Tennekes and Lumley 1972, 
Meneveau and Sreenivasan 1987, 1991), and considerable attention has been
devoted to multifractals in turbulence (Frisch and Parisi 1985, Mandelbrot
1989).

A working definition of a fractal is a structure which is generated by
sequential application of a scale-invariant rule on regularly decreasing
spatial scales. Simple fractals with constant (but non-integer) dimension are
invariant to changes in scale, and result from rules which produce a binary
distribution of the local density (say, either 0 or 1). Examples of these are
the Cantor set or the Sierpinski gasket (Mandelbrot 1989), in which segments of
a line, or portions of a plane, are simply removed without changing the
surrounding values. Their average density, as a function of scale $\epsilon$,
may be written as $\rho(\epsilon) = \rho_0 \epsilon^{-d}$ with dimension
$d=0.63$ (between a point and a line) and $d=1.52$ (between a line and a plane)
respectively.

In contrast, {\it multifractals} result from application of rules (or
probability distributions of rules) in which the local measure is {\it
changed}, while conserving the total measure, by unequal repartitioning of the
content of a bin into sub-bins of regularly decreasing spatial scale. For
example, it is easily seen that partitioning some quantity into the two halves
of a bin with unequal proportions (say 0.7 and 0.3) raises the mean density in
the first half and decreases it in the second. Repetition of this rule produces
some bins which become ever denser with decreasing scale, and others which
become ever less dense - with all combinations in between. The ensuing spatial
distribution has {\it no well defined local value} in the limit of diminishing
bin size; that is, the local values are spatially spiky, ``intermittent", or
``singular" (Meneveau and Sreenivasan 1987 present a short and  readable 
discussion of cascade processees and multifractals). The spatial distributions
of multifractals are, however, predictable in a statistical sense, using
Probability Distribution Functions (PDFs) which are derived directly from their
dimensions. For instance, {\it dissipation of turbulent kinetic energy}, which
occurs on the Kolmogorov spatial scale, is not spatially uniform but has the
spatial distribution of a mutifractal (Chhabra et al 1989).

In multifractals, the dimension {\it varies} with the value of the {\it
measure}. In our case, the measure is particle concentration factor $C$, which
is the ratio of the local particle volume density to its global average value.
The fractional probability $P_i$ of a particle lying in a bin
which contains $N_i$ particles out of $N_p$ total particles is defined as $P_i
= N_i/N_p \equiv \epsilon^{a_i}$, where $\epsilon \equiv$ bin size/domain size.
The scaling index $a$ can be viewed as a local dimension for $P$. 
The associated concentration factor $C_i \equiv { N_i/v_i \over N_p/v}$, where
$v_i$ is the volume of a bin and $v$ is the total volume of the domain; thus
$C_i = { N_i/N_p \over v_i/v } = (\epsilon^{a_i} / \epsilon^3) =
\epsilon^{a_i-3}$. Expressing the bin size, or scale, 
as some multiple $B$ of $\eta$, and
the domain size in units of the integral scale as $DL$, we find the domain
normalized bin size to be $\epsilon = B\eta/DL = \frac{B}{D}Re^{-3/4} \equiv
1/{\cal R}$, where we have also used the inertial range relationship $\eta = L
Re^{-3/4}$. Thus, $C = {\cal R}^{3-a}$, or
\begin{equation}
a = 3 - { {\rm ln} C \over {\rm ln}{\cal R}}.
\end{equation}
The normalized PDF for $a$ is usually written as a fractional volume $F_v(a) =
\rho(a) \epsilon^{3-f(a)}$, where the important function $f(a)$, called the
{\it singularity spectrum} (Halsey et al 1986, Chhabra et al 1989, Mandelbrot
1989), plays the role of a dimension for $F(a)$. The prefactor function
$\rho(a)$ is only weakly dependent on scale, and can be approximated as
$\sqrt{{\rm ln}{\cal R}}$ (Chhabra et al 1989). The function $f(a)$ is
discussed more in the next subsection.

We define the PDF $F_v(C)$ as the volume fraction occupied by bins having
concentration factor within $(C, C+dC)$, with $\int_{C_{min}}^{C_{max}}
F_v(C)dC = \int_{a_{max}}^{a_{min}} F(a)da \equiv 1$. Transforming variables
and their PDFs, we get

\begin{equation}
F_v(C) = F(a) \mid{ da \over dC} \mid = { \rho(a) {\cal R}^{f(a)-3} \over C
{\rm ln} {\cal R}}  \approx 
{ {\cal R}^{f(a)-3} \over C \sqrt{{\rm ln} {\cal R}}}.
\end{equation}

\begin{figure} 
\centerline{\psfig{figure=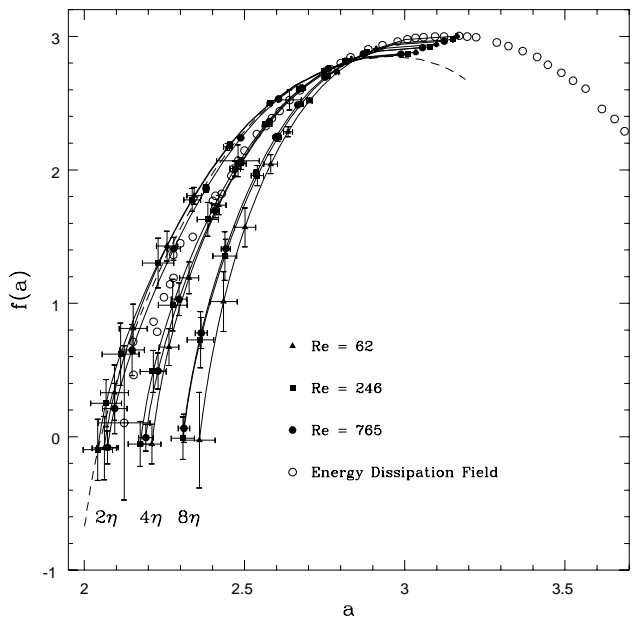,width=3.5 in,height=3.5 in}} 
\caption {Singularity spectra $f(a)$ for particle concentration (solid
symbols), for bin sizes of 2, 4, and 8 $\eta$, along with error bars
representing deviations across many snapshots. The similar looking singularity
spectrum for dissipation (open symbols) is from Meneveau and Sreenivasan (1987).
Also shown (dashed curve) is our annealed $f(a)$ for $2\eta$ binning, covering
$C$ values which occur less than once per computational domain (and thus
extends to negative values). Figure adapted from HCD99.}
\vspace{0.5 in}
\end{figure}

The fraction of {\it
particles} occupying bins within $(C, C+dC)$ is defined as $F_p(C) = C F_v(C)$,
and both $F_v(C)$ and $F_p(C)$ have cumulative versions $F_v(>C) =
\int_{C}^{\infty}F_v(C) dC$, and $F_p(>C) = \int_{C}^{\infty}F_p(C) dC$. For a
stationary, ergodic process, one expects that the fraction of {\it particles}
having $C$ at any given time ($F_p(C)$) is identical to the fraction of {\it
time} spent in regions of concentration $C$ by a typical particle, defined as
$F_t(C)$. Below, this is confirmed numerically.

\subsection{Numerical results at low $Re$:}
\subsubsection{PDFs for concentration}
Using the methodology of Chhabra et al (1989), we have shown that the spatial
distribution of optimally concentrated ($St_{\eta}=1$) particles is a
multifractal, and that its singularity spectrum $f(a)$ is invariant over more
than an order of magnitude from $Re=62$ to $Re=765$ (HCD99, who actually give
Taylor microscale Reynolds numbers $Re_{\lambda} = \sqrt{15 Re}$ = 40 to 140).
This $Re$-invariance for $f(a)$ is seen only when binning is done on some
fundamental flow-relative scale such as a multiple of the Kolmogorov scale
$\eta$. The function $f(a)$ is found by applying weighted binning techniques to
the distribution of interest (Chhabra et al 1989, HCD99), and depends on both
$D$ and $B$ - that is, on the number of integral scales sampled as well as on
the binning scale. Careful reassessment of different definitions of the
integral scale (see Hinze 1975, chapter 2) has led us to values slightly
smaller than those published by HCD99, implying that the computational domain
appropriate for our $f(a)$ is 9$L$ on a side. Our results for $f(a)$ relative
to this domain are shown in {\bf Figure 3} (HCD99). The sets with error bars
are ``quenched", or averaged over samples each $DL$ in extent. The fact that
there is no less than one cell in each sample found at maximum concentration
$C_{\rm max} = {\cal R}^{3-a_{\rm min}}$ results in $f(a_{\rm min})=0$ for
these sets. This can be seen by setting the product of either $F_v(a)$ or
$F_v(C)$ by the normalized domain volume ${\cal R}^3$ equal to unity, after
allowing for the fact that the $F_v$ are differential functions of their
arguments by multiplying them by $a_{\rm min}$ or $C_{\rm max}$ respectively.

Information about higher (less probable) concentrations than are typically
found in a volume $DL$ on a side must come from analyzing a large number of
realizations of each sample and determining an ``annealed" average, where
$f(a_{\rm min})<0$ ({\it eg.,} Chhabra and Sreenivasan 1991). The annealed
version (dashed curve in {\bf figure 3}), also binned by $2\eta$, is well fit
by the function $f(a) = -12.414 + 89.659/a -132.01/a^2$, for $a > a_{\rm min} =
$ 2.0. We have restricted ourselves to $2\eta$ binning for the present to
preserve good statistics; smaller binning scales sample a deeper cascade and
will generate large $C$ values with higher probability. The three sets of
$f(a)$ in {\bf Figure 3}, as binned over 2, 4, and 8 $\eta$ illustrate the
multifractal or ``singular" nature of the distribution. For any given $Re$,
smaller $a$ values correspond to larger $C$ values (equation 9), and larger
values of $f(a)$ imply larger $F_v$. The smaller $a_{\rm min}$, or larger
$f(a)$ for $a>a_{\rm min}$ seen for the smaller binning scales implies that,
averaged over a bin, smaller binning scales retain large $C$ far more commonly
than larger binning scales. That is, as the scale is reduced, no asymptotic or 
well-defined limiting local value is reached. Also shown in {\bf Figure 3} is 
the singularity spectrum for dissipation of turbulent kinetic energy (open 
symbols), which is scale-independent. The agreement between this spectrum and
that for particle density binned at the $2\eta$ scale is intriguing. The
deviation between our spectra (solid symbols and dashed line) and that for
dissipation seen toward the right hand side (large $\alpha$) is due to
incomplete sampling of very {\it low} particle density regions in our particle
density simulations because of memory limitations on the number of particles we
can follow. Dissipation, being a continuously varying function, is not subject
to this effect. For this reason, $F_v(>C)$ was less well defined than
$F_p(>C)$, which, by definition as a {\it particle} rather than {\it volume} 
fraction is always fully characterized. In any case, our primary interest is in
the zones of high concentration (small $\alpha$).

\begin{figure}
\centerline{\psfig{figure=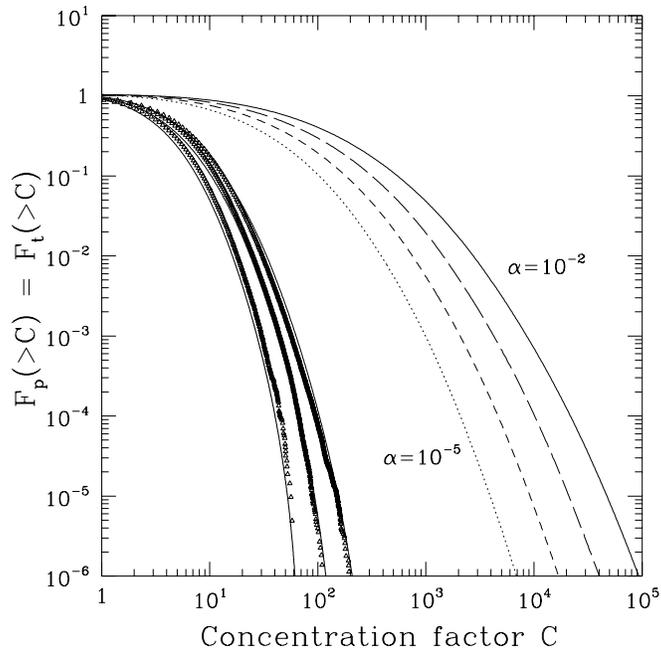,width=3.5 in ,height=3.5 in}}
\caption[title for caption]{Probability Distribution Functions (PDF's) for the
fraction of particles lying in regions with concentration factor greater than
$C$ ($F_p(>C)$), or, equivalently, the fraction of time spent by any particle in
such regions ($F_t(>C)$). The three sets of points are binned directly from our
numerical simulations; the associated curves are calculated from a single
averaged $f(a)$ obtained from all three values of $Re$ (dashed curve in {\bf
Figure 3}). As discussed in section 5.2, the curves without points use the {\it
same} $f(a)$ to predict PDF's at the larger $Re$ corresponding to four
plausible nebula $\alpha$ values: $10^{-2}$ (solid line), $10^{-3}$ (long
dashed line), $10^{-4}$ (short dashed line), and $10^{-5}$ (dotted line). }
\label{fig:pdfs}
\end{figure}

As a check on the method (subtle normalization issues are discussed in Chhabra
et al 1989), an average $f(a)$, obtained from our calculations at all three
$Re$ values was used to calculate $F_p(>C)$ at each of our three $Re$ values,
for comparison with the $F_p(>C)$ distributions determined directly from
numerical results (HCD99).  As shown in {\bf figure 4}, the single $f(a)$ does
quite well at predicting $F_p(>C)$ at all three $Re$, even though they all
subtended slightly different numbers of integral scales. For any $C$, $F_p(>C)$
increases with $Re$ and thus {\cal R}. This may be puzzling after only a quick
inspection of equation (10), since $f(a)-3$ is always negative ({\bf 
figure 3}). However, for a fixed $C$, 
$a(C)$ increases with ${\cal R}$ (equation 9),
and thus $f(a)$ also increases ({\bf figure 3}), so the exponent in equation
(10) becomes {\it less} negative. Since any function is more sensitive to 
its exponent (here $f(a)-3$) than to its base (here ${\cal R}$), 
equation (10) implies $F_v(>C)$ and thus $F_p(>C)$
increase with ${\cal R}$. 

\subsubsection{Particle time histories}
The particle time histories are of interest in their own right, and illustrate
how particles experience a fluctuating background concentration as they wander
through the fluid. {\bf Figure 5} illustrates histories for several randomly
chosen particles. The particles are moving at roughly constant space velocity
(approximately the velocity $V_L$ of the largest eddies, since they are trapped
to nearly all eddies), and repeatedly encounter zones of different density with
little noticeable effect on their velocity (their stopping time is much longer
than the clump transit time $\eta/V_L$). In these simulations, where no particle
interactions are computed, the particles pass through the dense zones and
continue their evolution. The dense zones {\it per se} persist for times much
longer than the passage time of a single particle (CDH96, She et al 1990). The
more numerous, lower density zones are encountered more frequently, and the
rare, very high density zones less frequently. This time history, essentially a
(convoluted) 1D path through the computational volume, has the same
``intermittent" or spiky structure as seen for dissipation (see, eg., Meneveau
and Sreenivasan 1987; Chhabra et al 1989).

\begin{figure}                                                             
\centerline{\psfig{figure=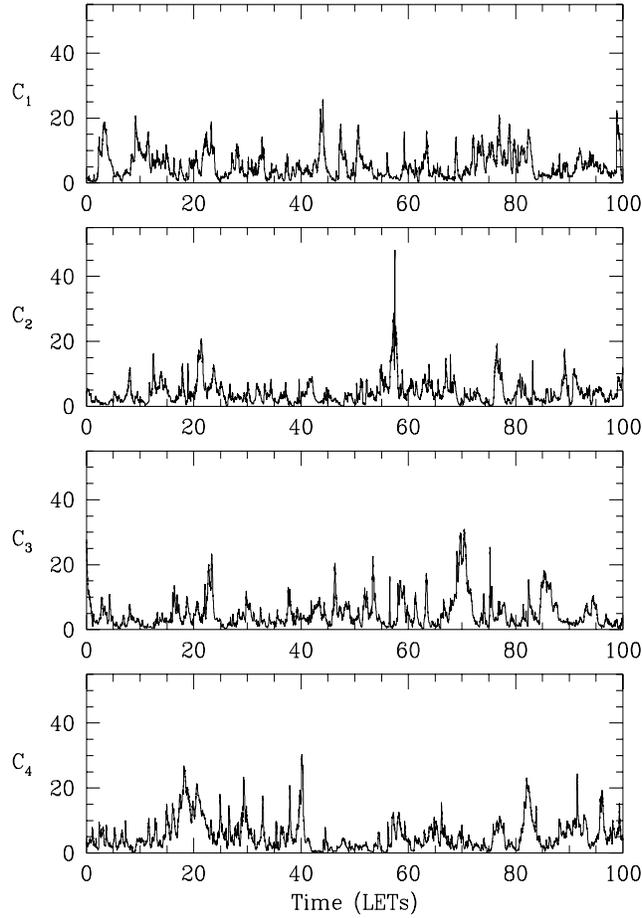,width=4.5 in,height=5.0 in}} 
\caption{Ambient concentration encountered by four different particles moving
in 3D turbulence, as a function of time (measured in large eddy turnover times
LET = $\omega(L)^{-1}$), without feedback onto the gas. These may also be
regarded as a longer history for a single particle. Note how denser regions are
encountered less frequently. }
\vspace{0.5 in}                                                         
\end{figure}

\begin{figure}
\centerline{\psfig{figure=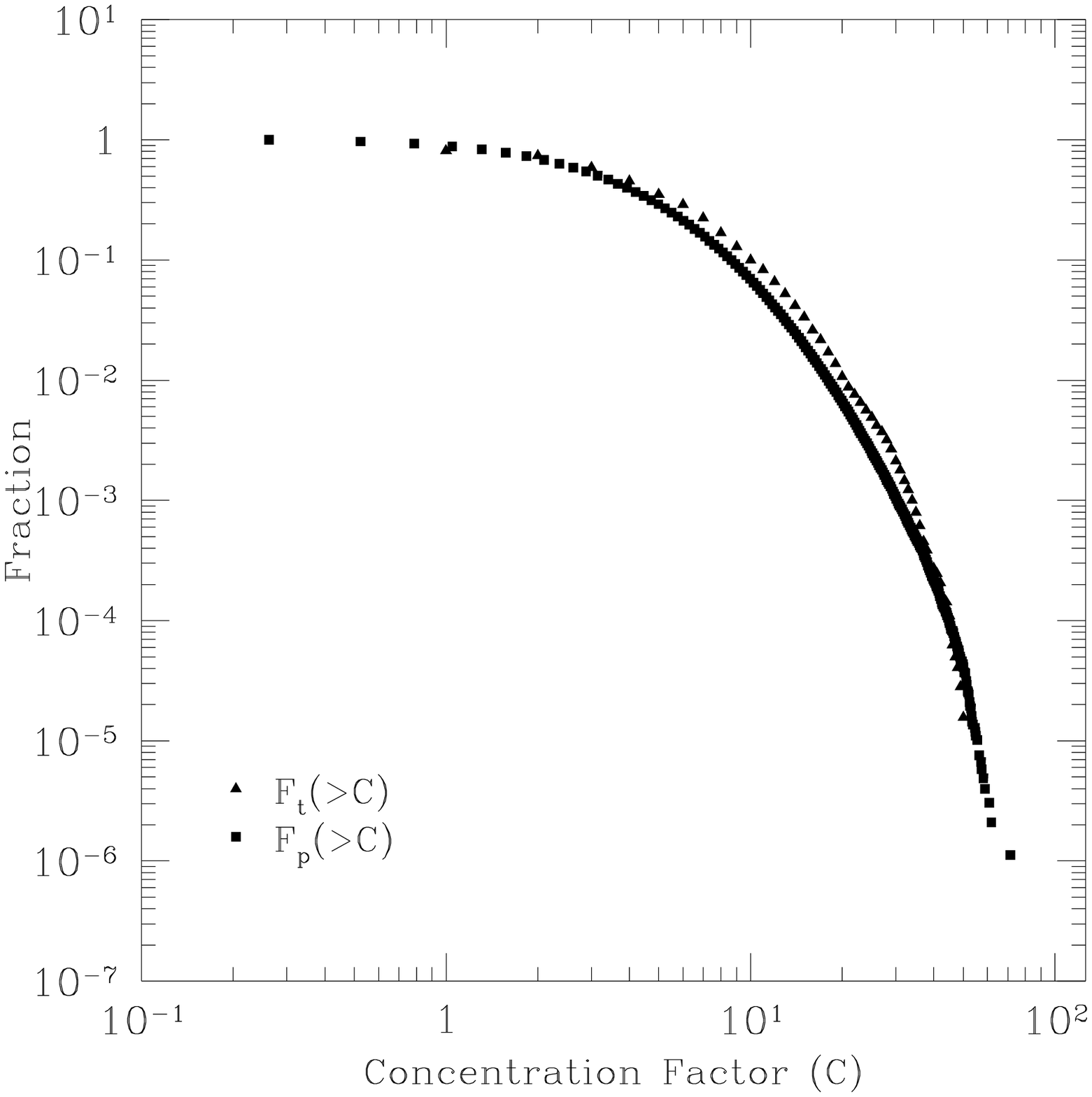,width=3.5 in,height=3.5 in}} 
\caption{Comparison of $F_p(>C)$ from binned data in several snapshots
(squares, temporal realizations) containing all particles (as in Figure 4), with
$F_t(>C)$ (triangles), calculated from extended time histories of 16 different
particles (as in Figure 5).}
\vspace{0.5 in}
\end{figure}
From simulations such as these, we have validated the ergodic assumption that
the fraction $F_p(>C)$ (spatially averaged over all particles at several
snapshots in time) is equal to the fraction of {\it time} $F_t(>C)$ spent by a
{\it given} particle in regions denser than $C$ (temporally averaged over
extended trajectories for a few particles). The comparison is shown in {\bf
figure 6}. A ``random walk" calculation does {\it not} accurately
reflect this behavior; the particles are not moving randomly through the volume
and encountering dense zones with probability given by their volume fraction;
rather, particle trajectories are ``attracted" to the dense zones, or, probably 
more physically, repelled from the sparse, complementary eddy zones. This
behavior is correctly captured using $F_p(C)$.

\subsection{Predictions for nebula $Re$:}
The properties of multifractals make it much easier for us to predict nebula 
properties than merely making extrapolations from $Re$ values we have studied 
numerically. In the multifractal context, equation (10), and the $Re$-
independence of $f(a)$, suggest identification of $f(a)$ with the
level-independent ``rule", and ${\cal R} \propto L/\eta = Re^{3/4}$ with the
number of steps in the cascade. As $Re$ increases, the inertial range (or
number of eddy bifurcations) between $L$ and $\eta$ also increases. The cascade
model described by Meneveau and Sreenivasan (1987), mentioned at the beginning
of this section, is a specific case of the binomial cascade discussed by
Mandelbrot (1989); all provide insight as to how higher $C$ result from a
deeper cascade with more steps. Mathematically, equations (9-10) show how high
$C$ values become more likely. If $f(a)$ is indeed a $Re$-independent,
universal function for optimally concentrated ($St_{\eta}=1$) particles, we can
then predict PDF's for any $Re$ ({\it i.e.}, any given nebula $\alpha$) 
using equation (10).

In addition to our own experiments over an order of magnitude in $Re$, several
other lines of argument support the assumption of a level-independent and
$Re$-independent ``rule". Dissipation of $k$ and particle concentration are
physically connected through their mutual preference for the Kolmogorov scale;
also, the shapes of the singularity spectra for dissipation, and for particle
concentration binned at close to the Kolmogorov scale, are similar ({\bf 
figure 3}; HCD99). The
singularity spectrum of dissipation has already been connected to the turbulent
cascade process (Meneveau and Sreenivasan 1987). Thus, turbulent concentration
is probably also closely related to the turbulent cascade process. The
turbulent cascade is known to have $Re$-independent properties in the inertial
range. For instance, the $f(a)$ for dissipation has been shown to be
$Re$-independent - from numerical work at $Re \sim 100$, including our own,
through laboratory experiments with $Re \sim 10^4$, to experimental studies of
the atmospheric boundary layer with $Re \sim 10^6$ (Chhabra et al 1989,
Hosogawa and Yamamoto 1990). Furthermore, it has been directly demonstrated
from analysis of observations of dissipation in large $Re$ turbulence that the
probability distribution of ``partition factors" or ``multipliers" is
independent of level within the inertial range (Sreenivasan and Stolovitsky
1995). Based on these arguments, we believe and presume that the particle
concentration singularity spectrum $f(a)$ remains $Re$-independent in turbulent
cascades with far larger values of $Re$ than those of our numerical
experiments. Given this invariance, we can predict nebula conditions more
confidently than from extrapolation alone, as was done by CDH96. We obtain
PDF's at $Re$ which are much larger than directly accessible values by
containing the $Re$-dependence purely within ${\cal R}= DL/B\eta =
(D/B)Re^{3/4}$. As argued at the beginning of this section, the $f(a)$ is
associated with a certain $D$ and $B$, which we interpret as being the $D
\approx 9$ and $B$=2 implicit in our $f(a)$ (section 5.1.1). The nebula ${\cal
R}$ is then only dependent on $Re$ and easily determined for any turbulent
$\alpha$ as ${\cal R}= 1 /\epsilon = D L / B \eta = (D/B) Re^{3/4} \approx 4.5
Re^{3/4}$. Recalling $\nu_m = m_{H_2} c / 2 \rho_g \sigma_{H_2}$ (section 3),
\begin{equation} 
{\cal R} = 4.5 \left({ \alpha c H \over \nu_m} \right)^{3/4}
= 4.5 \left( {\alpha \Sigma \over m_{H_2}/\sigma_{H_2} } \right)^{3/4} = 
7 \times 10^5  \left( {\Sigma \over 430 {\rm g}{\rm cm}^{-2}}\right)^{3/4}  
\left( {\alpha \over 10^{-4} } \right)^{3/4}.
\end{equation}
Also shown in {\bf figure 4} are predictions of $F_p(>C) = F_t(>C)$ for four
values of nebula $\alpha$, using equations (10) and (11). Based on these
predictions, $St_{\eta}=1$ particles spend about 1 - 10\% of their time in
regions with $C > 10^3$ under nebula conditions at different $\alpha$. All
curves in {\bf figure 4} assume a minimum mass nebula ($\Sigma = 430$ g cm$^2$
at 2.5 AU), but mass enhancement by some factor {\cal F} would play a role
similar to $\alpha$ (see also section 3).

\section{Implications and discussion}

\subsection{Regions of moderately high density}
For concentration factors leading to $\rho_{\rm ch}/\rho_g$ at least as large
as unity, particle mass loading might start to affect the ``partition rules"
and change the statistics of the cascade. Nevertheless, several interesting
effects can result from concentrations less than this possible limit. If the
region in question  has an average mass density in solids of $\rho_p = f_{\rm
sol}\rho_g = 5-8 \times 10^{-3}\rho_g$ (at, say, 400$^{\circ}$K; Pollack et al
1994), and some fraction $f_{\rm ch} < 1$ of this amount is in chondrule-sized
particles (recalling that other particle sizes are not susceptible to TC), it
is clear that $C$ as large as $(f_{\rm sol}f_{\rm ch})^{-1}$, at least several
hundred (for $f_{\rm ch}=1$), remains free of mass loading concerns. If $f_{\rm
ch}<<1$, far larger $C$ are allowed. The point is that, depending on $f_{\rm
sol}f_{\rm ch}$, $C$ can get quite large without violating the mass loading
caveat.

While most of the {\it volume} of the nebula is characterized by $C < 100$,
{\bf Figure 4} shows that chondrules in the nebula spend a significant fraction
of their {\it time} residing in regions where the particle density is enhanced
by a factor of $C > 100$. These results might help us understand why some
chondrule types seem to have been heated under unusually oxidizing conditions
({\it eg.,} Rubin et al 1989), which has been attributed to vaporization of a
large ambient density of solids - enhanced by several orders of magnitude over
solar - as part of the chondrule formation process (note, however, that it is
not universally accepted that the chondrule oxidation state information
necessarily implies a large background abundance of oxygen; Grossman 1989).

Several chondrule melting processes - shock waves (Hood and Kring 1996,
Connolly and Love 1998) or lightning bolts (Desch and Cuzzi 2000, Desch 2000)
are envisioned to occur ubiquitously within the region of the nebula in which we
propose TC operates to sort the chondrules after their formation. In fact,
Desch and Cuzzi (2000) showed that TC itself is an enabling factor in
generation of nebula lightning. They found that the optimal conditions for
energetic lightning bolts are found in 1000-km regions with $C \sim 100$ -
especially if the nebula were denser than ``minimum mass" - ${\cal F} \sim 10$.
Either lightning or shock waves - heating dense zones where the background
density of solids is large ({\it cf.} Hood and Horanyi 1993, Hood and Kring
1996, Connolly and Love 1998) - could elevate the local Oxygen abundance by
evaporating some part of the solids. However, TC does not concentrate ``fine
dust" - only chondrule-sized particles - so significant local volatilization of
some chondrules, or the surfaces thereof, would be implied. Whether this is
consistent with mineralogical signatures which are seen in the survivors would
be useful to address in the future.

\subsection{Encounters with very dense regions}
The scenario of CDH96 suggested in a qualitative way that TC could lead to
stable, if low-density, clusters or clumps of chondrules as direct precursors
of planetesimals. We delineate the logic of this argument, using the PDF's
derived in section 5, and then discuss the difficulties with the original
scenario.

The PDF's can be combined with the particle velocity through space $V_p$ to
calculate the ``encounter time" $T_{\rm enc}$ of a chondrule with a region of
arbitrary $C$, using a duty cycle argument. The fraction of time spent by a 
typical chondrule-size particle in clumps with concentration greater than $C$, 
$F_t(>C) \approx { t_{\rm in}(>C)
\over t_{\rm out}} \approx { t_{\rm in}(>C) \over T_{\rm enc}}$, where $ t_{\rm
in}(>C) << T_{\rm enc}$ is the time spent traversing a bin with concentration
greater than $C$ (section 5.1.2). Thus, for bins of dimension $2\eta$ and
particle velocity $V_p$, $ t_{\rm in}(>C) = 2 \eta/V_p$, so $T_{\rm enc}
\approx { t_{\rm in}(>C) \over F_t(>C)} = { t_{\rm in}(>C) \over F_p(>C)} = { 2
\eta \over V_p}{1 \over F_p(>C)}$, and the encounter rate is ${1 \over T_{\rm
enc}} = { V_p \over 2 \eta }F_p(>C)$.  We have verified numerically that, as
expected for particles with stopping times $t_s$ much shorter than the overturn
time of the largest eddies $\Omega_0^{-1}$, $V_p$ is nearly identical to the
typical turbulent gas velocity $\sqrt{2k} \approx c \sqrt{\alpha}$ (V{\"o}lk et
al 1980, Markievicz et al 1991).

We may calculate the encounter rate (and time) with a cloud so dense that a
particle becomes entrapped with its neighbors and possibly 
removed from further free
circulation. Normally, as seen in {\bf figure 5}, particles traverse dense
regions without incident, because their gas drag stopping time is longer than
their transit time (CDH96). An entrapment threshold occurs if interparticle
collisions can prevent particles from passing through a cloud; this implies a
critical cloud optical depth $\tau_{\rm coll}$ of unity, defining a critical
$C_{\rm coll}$. For small, dense $2\eta$-sized clumps, $\tau_{\rm coll} = 1 =
2\eta C_{\rm coll} (\rho_{\rm ch}/m) \pi (2r)^2$, where $\rho_{\rm ch} = f_{\rm
sol} f_{\rm ch} \rho_g$ is the average (unconcentrated) chondrule mass density,
$f_{\rm sol} \approx 5 \times 10^{-3}$ at 400K is the fractional mass in
solids, $f_{\rm ch}$ is the fraction of {\it solids} in chondrules, and $m$ and
$r$ are chondrule mass and radius. Then
\begin{equation}
C_{\rm coll} \approx 33 r \rho_s / \eta \rho_g f_{\rm ch} 
\approx {4 \times 10^5 \over f_{\rm ch}}  
\left( {10^{-4} {\cal F} \over \alpha}\right)^{1/4}.
\end{equation} 
In the final expression above, we have substituted the first expression of
equation (8) for the product $r \rho_s$, since the value of this product 
for optimally concentrated particles is constrained by nebula conditions
(section 3). We substituted the general expression for $\eta = L Re^{3/4} =
(H/\alpha)^{1/4}(\nu_m/c)^{3/4}$ ($\sim$ 0.5km $(10^{-4}/\alpha)^{1/4}$), and
assumed other nominal parameters at 2.5 AU ($\Sigma = 430$ g cm$^{-2}$, $c=1.5
\times 10^5$ cm/sec).

For nominal, minimum mass nebula parameters, assuming $f_{\rm ch} \approx 1$,
and using the PDF's we have in hand (binned at $2 \eta$ scales), we find that
$T_{\rm enc}$ varies between $10^3$ years for $\alpha = 10^{-2}$ to $10^8$
years for $\alpha = 10^{-4}$. The encounter times are sensitive to nebula
parameters adopted and would decrease for smaller binning ({\it eg.}, by
$B=1$), or enhanced nebula densities (${\cal F} > 1$). This quantifies the 
length of time chondrules spend freely wandering before encountering a region 
in which they are certain to undergo collisions. We suggested earlier that 
such an encounter at $T_{\rm enc}(C_{\rm coll})$ removes the chondrule from
further circulation by entrapping it with others in the dense cluster.
Alternately, such dense zones might merely provide a large increase in the
collisional
aggregation rate of optimally sized particles. In fact, if chondrules or groups
of chondrules undergo collisional aggregation at low relative velocities (and
particles of comparable size always have low relative velocities in turbulence;
Weidenschilling and Cuzzi 1993), they may form fractal aggregates with mass
proportional to radius {\it squared} (Weidenschilling and Cuzzi 1993, Beckwith
et al 2000). The average density of such aggregates is inversely proportional
to their bounding ``radius", so they retain the same stopping time as their
individual components (individual chondrules), and can continue to participate
in TC.

However, before this line of thought can be pursued much further, mass loading
(discussed in the next section) must be assessed. Depending on their linear
extent, the optically thick clouds described above can reach a mass density
orders of magnitude larger than that of the gas, which would probably
invalidate our assumption of no feedback. In general, for a clump of mass
density $\rho_{\rm ch}$ composed of chondrules of radius $r$ and density
$\rho_s$,
\begin{equation}
\rho_{\rm ch} = {r \rho_s \tau_{\rm coll} \over 3 B\eta};\hspace{0.1in}{\rm 
or}\hspace{0.1in} {\rho_{\rm ch}\over\rho_g} = {r \rho_s \over 3 B\eta \rho_g}
\hspace{0.2 in} {\rm for} \hspace{0.2 in} \tau_{\rm coll}=1.
\end{equation}
In deriving equation (13) we have expressed the linear size of a cluster as
$B\eta$. For $\eta \sim 0.5$ km $(10^{-4}/\alpha)^{1/4}$ and $\rho_g= 1.1
\times 10^{-10}$ g cm$^{-3}$, the particle density exceeds $10 \rho_g$ for
$\tau_{\rm coll}=1$ clusters smaller than $(60-600/{\cal F})$ km - about
$(100-1000/{\cal F})\eta$ for $r \rho_s$ in the normal range for chondrules
$\approx 0.02 - 0.02$ (figure 1).

\subsection{Limitations due to high mass loading:}

The cascade process model is only valid as long as no new physics emerges at
some step in the cascade to change the partition factors, as represented
globally by $f(a)$, applicable to subsequent steps in the cascade. However,
application of the expressions above for optically thick clusters of scale
$2\eta$ implies $\rho_{\rm ch}/\rho_g \sim 3000$. While these high
concentrations literally relate to regions comparable to $\eta$ in size, where
there is no turbulence to damp, the cascade that is needed to {\it produce}
such a cluster must have extended to larger sizes in the surrounding
``penumbra" where the particle density, while lower, might still be large
enough to damp turbulent motions. We have made some preliminary calculations of
$f(a)$ from low-$Re$ numerical simulations having mass loading $\rho_p/\rho_g
\sim 1$. Even here, turbulent concentration persists (to $\rho_p \sim 30
\rho_g$), but the $f(a)$ is altered in
the sense that high $C$ values have a lower probability. For comparison,
unloaded simulations at this $Re$ result in $C_{\rm max}\approx 60$. The
magnitude of the mass loading effect is approximately equal and opposite to
that of decreasing the bin size from $2\eta$ to $\eta$. Clearly, this effect
must be better quantified before more specific predictions of $T_{\rm enc}$ and
chondrule accumulation timescales can be made. Even for small $\alpha \sim
10^{-3}$, the mass loading regime of questionable validity is not a large one
(note that equation (13) above showed that regions of size $300\eta$, or
100-300 km, are probably within the range of validity); however, it is
certainly an interesting one. If we further restrict our attention to optimally
concentrated particles by combining equations (7 or 8) and (13), we obtain
\begin{equation} {\rho_{\rm ch}\over\rho_g} = \left({ m_{H_2} \over
4 \sqrt{2} \sigma_{H_2}} \right)^{1/2} \left({ 1 \over 3 \sqrt{2} B \eta }
\right) \left({ H \over \alpha \rho_g} \right)^{1/2}
\end{equation} 
which shows that
regions which are denser and/or have more intense turbulence (higher $\alpha$),
such as may have characterized the very early evolutionary stages of the
nebula, are less prone to mass loading difficulties. That is, they provide
$\tau_{\rm coll}=1$ at lower $\rho_{\rm ch}/\rho_g$. {\bf Figure 7} shows that
conditions at 1AU (gas density $\rho_g \sim 10^{-9}$ g cm$^{-3}$), with $\alpha
\sim 10^{-2}-10^{-1}$, provide $\tau_{\rm coll}=1$ with $\rho_p/\rho_g < 100$.
Full treatment of a wider parameter space, allowance for mass loading and other
binning scales $B$, and connection to hypotheses for chondrule formation (Desch
and Cuzzi 2000) and dust rimming (Cuzzi et al 1999, Morfill and Durisen 1998)
which are tied to TC, will be treated in subsequent publications.

\begin{figure}
\centerline{\psfig{figure=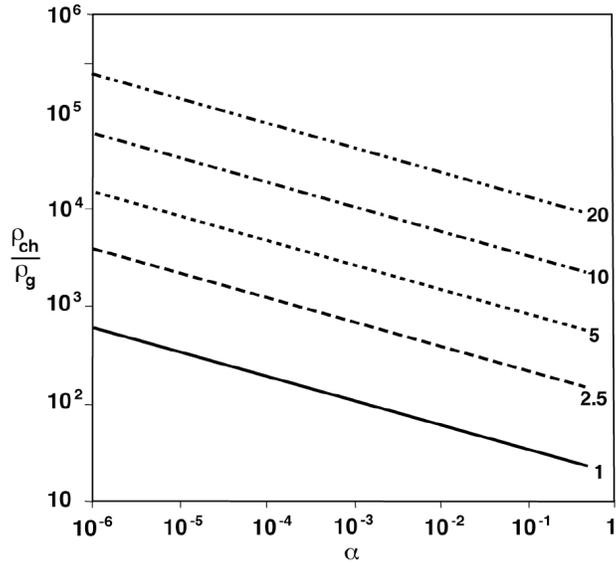,width=3.5in,height=3.5in}}
\caption{Variation of $\rho_p/\rho_g$ with $\alpha$ (for cluster optical depth 
unity) for the same nebula regions as shown in figure 1. Regions with more
severe turbulence (larger $\alpha$) select smaller particles for concentration
which, having larger area per unit mass, can be concentrated to $\tau_{\rm
coll}$ without reaching such large mass loading ratios as under more ``nominal"
conditions as shown in figure 1.}
\vspace{0.5 in}
\end{figure}

\section{Discussion:} Turbulent concentration promises to provide another tool
for our planetesimal construction toolkit. Nominal nebula conditions, including
weak turbulence, lead to concentration factors large enough to concentrate
chondrule-sized particles to densities greatly exceeding the nominal nebula
average. The presence of such dense regions might help us understand some 
aspects of chondrule formation. Generalization of the physics shows that fluffy
aggregates would also be subject to turbulent concentration, but in nebula
regions where the gas density is lower and/or the turbulent intensity higher
than expected for the terrestrial planet region. If TC is a key aspect of 
the primary accretion of planetesimals, any chondrules {\it formed} or {\it 
transported} to low density, outer solar system regimes might only have 
escaped incorporation into planetesimals to a greater degree than in the 
terrestrial planet region. The size distributions produced by turbulent
concentration are in very good agreement with those found in primitive
chondrites; however, the fact that most primitive meteorites contain ample
evidence for abrasion, fragmentation, and other mechanical processes which may
well have continued long after the nebula gas vanished and aerodynamic
processes became irrelevant, warns us that we should not expect any single
process to explain all properties of even ``primitive" meteorites.

The spatial distributions of concentrated chondrules under nebula conditions 
can be predicted using the multifractal properties of turbulent concentration. 
Various timescales of interest can then be estimated, including the timescale 
in which a chondrule encounters a cluster so dense that it is entrapped. 
However, preliminary estimates of timescales for simple accumulation by
entrapment in unusually dense clusters are comfortably shorter than nebula
lifetimes only for higher $\alpha$ values than those which best explain
chondrule size-density products. 

Mass loading is expected to change the nature of the process at some point in 
the cascade where the concentrated particle density becomes comparable to, or 
greater than, that of the gas. Depending on the mass fraction in solid 
``chondrules", this might imply concentration factors as small as $10^2$ - or 
possibly far larger. Mass loading has not yet been quantitatively treated, but 
it seems it will inevitably increase the encounter timescales by decreasing the
abundance of very dense clusters. Perhaps turbulent concentration only augments
collisional accumulation for optimally sized particles.

The evolution of dense clusters in the presence of mass loading and the
vertical component of solar gravity deserves study (eg. Wang and Maxey 1993).
Do clusters retain their identity or disperse? Would they be compacted or
dispersed by settling, by collisions with other dense clusters {\it en route}
to, or within the midplane, or by shock waves in the nebula? Perhaps only
clusters formed near the midplane avoid the dispersion due to settling. 

{\bf Acknowledgements:} This research was jointly supported by NASA's Planetary
Geology and Geophysics Program and Origins of Solar Systems Program. We thank
John Eaton, Kyle Squires, Pat Cassen, Steve Desch, Peter Goldreich, and K. R.
Sreenivasan for helpful conversations. We also thank numerous meteoriticist
colleagues for their interest and their patient tutorials - especially Ted
Bunch, William Skinner, John Wasson, Harold Connolly, Alan Rubin, and Dotty
Woolum. The extensive computations on which much of this research is based were
made possible by the capabilities of the National Aerodynamical Simulator (NAS)
Program at Ames Research Center; we thank Eugene Tu for extra discretionary
time. We thank the Antarctic Meteorite Working Group for making samples of ALH
85033 and other meteorites available. This research has made use of NASA's
Astrophysics Data System Abstract Service.

\setlength{\parindent}{-0.2in}
\section*{Bibliography}

Balbus, S. A. and J. F. Hawley (1991) Ap. J. 376, 214-233.

Balbus, S. A. and J. F. Hawley (1998) Revs. Mod. Phys., 70, 1

Balbus, S. A., J. F. Hawley, and J. M. Stone (1996) Ap. J. 467, 76-86

Barge, P. and J. Sommeria (1995) Astron. Astrophys. 295, L1-L4

Bell, K. R., P. Cassen, H. Klahr, and Th. Henning (1997); Ap. J. 486, 372-387

Boss, A. (1996) in ``Chondrules and the Protoplanetary Disk", edited by R. H.
Hewins, R. H. Jones, and E. R. D. Scott, Cambridge University Press, Cambridge

Bracco, A., P. H. Chavanis, A. Provenzale, and E. A. Spiegel (1999) Phys. 
Fluids, in press

Cabot, W., V. Canuto, and J. B. Pollack (1987) Icarus 69, 387-422

Cameron, A. G. W. (1978); Moon and Planets 18, 5

Chhabra and Sreenivasan (1991) Phys. Rev. A. 43, 1114-1117

Chhabra, A. B., Meneveau. C., Jensen, V. R. and Sreenivasan K. R. (1989),
Physical Review A 40, 5284-5294

Chokshi, A., A. G. G. M. Tielens, and D. Hollenbach (1993) Ap. J. 407, 806-819

Connolly, H. C. and S. Love (1998) Science 280, 62-67

Cuzzi, J. N., A. R. Dobrovolskis, and J. M. Champney (1993) Icarus, 106, 
102-134

Cuzzi, J. N., A. R. Dobrovolskis, and R. C. Hogan (1996)  in ``Chondrules and
the Protoplanetary Disk", edited by R. H. Hewins, R. H. Jones, and E. R. D.
Scott, Cambridge University Press, Cambridge; (CDH96)

Cuzzi, J. N. et al (1998) 29th LPSC Abstracts, on CD-ROM (Lunar and Planetary 
Institute).

Cuzzi, J. N., R. C. Hogan, and J. M. Paque (1999) 30th LPSC Abstracts, on 
CD-ROM (Lunar and Planetary Institute).

Desch, S. J. (2000) Submitted to Ap. J. 

Desch, S. J. and J. N. Cuzzi (2000) Icarus, in press (Protostars and Planets IV
special issue)

Dodd, R.T. (1976) E. P. S. L. 30, 281-291

Dominik, C. and A. G. G. M. Tielens (1997)

Dubrulle, B. (1993), Icarus 106, 59-76

Dubrulle, B., M. Sterzik, and G. E. Morfill (1995) Icarus, 114, 237-246

Eaton, J. K. and J. R. Fessler, (1994) Int. J. Multiphase Flow 20, Suppl.,
169-209.

Eisenhour, D. (1996) Meteoritics, 31, 243-248

Fessler, J. R., J. D. Kulick, and J. K. Eaton (1994) Phys. Fluids 6, 3742-3749

Frisch, U. and Parisi, G. (1985) in ``Turbulence and predictability in
Geophysical Fluid Dynamics and Climate dynamics", M. Ghil, ed. (N.
Holland-Amsterdam)

Gammie, C. (1996) Astrophys. J. 457, 355      

Goldman, I. and A. Wandel (1994) Astrophys. J. 443, 187-198

Grossman, J., A. E. Rubin, H. Nagahara, and E. A. King (1989) in ``Meteorites
and the Early Solar System", 619-659, J. Kerridge, ed; Univ. of Arizona Press

Grossman, J. (1989) in ``Meteorites and the Early Solar System", 619-659, J.
Kerridge, ed; Univ. of Arizona Press

Halsey, T. C., M. H. Jensen, L. P. Kadanoff, I. Procaccia, and B. I.
Shraiman (1986); Phys. Rev. A. 33, 1141-1151

Hartmann, L., N. Calvet, E. Gullbring, and P. D'Alessio (1998); Accretion
and the evolution of T Tauri disks. Astrophys. J. 495, 385-400

Hayashi, C. (1981) Prog. Theor. Phys. Suppl. 70, 35-53

Hewins, R. (1997) Ann. Revs. Earth and Planetary Sci., p 61

Hewins, R.,  R. H. Jones, and E. R. D. Scott (1996) ``Chondrules and the
Protoplanetary Disk", Cambridge University Press, Cambridge

Hinze, J. O. (1975) Turbulence, 2nd Ed. McGraw-Hill, New York,  Chapter 3

Hogan, R. C., and J. N. Cuzzi (2000) in preparation

Hogan, R. C., Cuzzi, J. N., and Dobrovolskis, A. R. (1999) Phys Rev E, 60, 
1674-1680 (HCD99)

Hood, L. and D. Kring (1996) in ``Chondrules and the Protoplanetary Disk",
edited by R. H. Hewins, R. H. Jones, and E. R. D. Scott, Cambridge University
Press, Cambridge

Hood, L. and M. Horanyi (1993) Icarus, 106, 179-189

Hosogawa, I. and K. Yamamoto (1990) Phys. Fluids A, 2, 889-892

Hughes, D. W. (1978) EPSL, 38, 391-400

Brearley, A. J. and R. H. Jones (1999) in ``Planetary Materials", Revs. in 
Mineralogy, 36, Ch. 3

Jones, R. et al (2000) in ``Protostars and Planets IV", V. Mannings, A. Boss,
and S. Russell, eds; Univ. of Arizona Press, in press.

Kato and Yoshizawa (1997) Pub. Ast. Soc. Jap. 49, 213-220

Kennard, E. H. (1938) Kinetic theory of gases; McGraw-Hill, New York (1938)

Keubler, K., H. McSween, W. D. Carlson, and D. Hirsch (1999) Icarus, 141, 
96-106

Klahr, H. (2000a) in Proceedings of the meeting ``Two decades of numerical 
astrophysics"; in press

Klahr, H. (2000b) in Proceedings of the meeting ``Disks, planetesimals and 
planets", Tenerife

Klahr, H. and Th. Henning (1997) Icarus, 128, 213-229

Lin, D. N. C. and J. Papaloizou (1985) in ``Protostars and Planets II"; D. C. 
Black and M. S. Matthews, eds, Univ. of Arizona Press

Lynden-Bell D. and J. E. Pringle (1974) MNRAS 168, 603-637

MacPherson, G. J., D. A. Wark, and J. T. Armstrong (1989) in ``Meteorites and 
the Early Solar System", 746-807, J. Kerridge, ed; Univ. of Arizona Press

MacPherson, G. J., A. M. Davis, and E. K. Zinner (1995) Meteoritics, 24, 297

Mandelbrot, B. (1989) Pure and Appl. Geophys, 131, 5-42

Markievicz, W. J., H. Mizuno, and H. J. V{\"o}lk (1991) Astron. Astrophys. 242,
286-289

Maxey, M. R. (1987) J. Fluid Mech. 174, 441-465

Meneveau, C. and Sreenivasan, K. R. (1987) Phys. Rev. Lett. 59, 1424-1427

Meneveau, C. and Sreenivasan, K. R. (1991) J. Fluid Mech. 224, 429-484

Metzler, K., A. Bischoff, and D. St{\"o}ffler (1992) Geochim. Cosmochim. Acta 
56, 2873-2897

Morfill, G. E. and R. H. Durisen (1998) Icarus 134, 180-184

Paque, J. and J. N. Cuzzi (1997) 28th LPSC Abstracts

Paque, J. and J. N. Cuzzi (2000) in preparation

Pollack, J. B, D. Hollenbach, S. Beckwith, D. Simonelli, T. Roush, and W. Fong
(1994) Astrophys. J. 421, 615-639

Prinn, R. J. (1990) Ap. J. 348, 725-729

Richard, D. and J.-P. Zahn (1999) Astron. Astrophys. 347, 734-738

Rubin, A. E., B. C. Fegley, and R. Brett (1989) in ``Meteorites and the Early
Solar System", J. Kerridge and M. S. Matthews, eds; Univ. of Arizona Press; p. 
488-511

Rubin, A. and J. Grossman (1987) Meteoritics 22, 237-251

Shakura, N. I. and R. A. Sunyaev (1973) Astron. Astrophys. 24, 337-355

She, Z.-S., E. Jackson, and S. A. Orszag (1990) Nature, 344, 226

Skinner W. R. and J. M. Leenhouts (1991) 24th L.P.S.C. Abstracts, 1315-1316

Squires, K. and J. K. Eaton (1990) Phys. Fluids A2, 1191-1203

Squires, K. and J. K. Eaton (1991) Phys. Fluids A3, 1169-1178

Sreenivasan, K. R.  G. Stolovitsky (1995) J. Stat. Physics 78, 311-333

Stone, J. M. and S. A. Balbus (1996) Ap. J. 464, 364-372

Supulver, K. and D. N. C. Lin (1999) Icarus, submitted

Tanga, P., A. Babiano, B. Dubrulle, and A. Provenzale (1996) Icarus 121, 
158-170

Tennekes, H., and Lumley, J. L. (1972) A First Course in Turbulence, MIT Press,
Cambridge Mass.

V\"{o}lk, H. J., F. C. Jones, G. E. Morfill, and S. R{\"o}ser (1980)
Astron. Astrophys. 85, 316-325

Wang, L. and M. R. Maxey (1993) J. Flu. Mech. 256, 27-68

Wasson, J. A. (1996) in ``Chondrules and the Protoplanetary Disk", edited by R.
H. Hewins, R. H. Jones, and E. R. D. Scott, Cambridge University Press,
Cambridge

Weidenschilling, S. J. (1977) Mon. Not. Roy. Ast. Soc. 180, 57-70

Weidenschilling, S. J. and J. N. Cuzzi (1993) in ``Protostars and Planets
III", E. Levy and J. Lunine, eds, Univ. of Az. Press

Wood, J. A. (1998) Astrophys. J. 503, L101-L104

\end{document}